\documentclass[twocolumn]{aastex631}
\usepackage{amsmath}

\begin{document}

\title{Green peas in the southern sky: Broad-band color selection and spectroscopic follow up}

\correspondingauthor{Hyunjin Shim}
\email{hjshim@knu.ac.kr}

\author[0009-0007-8971-7962]{Yejin Jeong}
\affiliation{Department of Earth Science Education, Kyungpook National University, 80 Daehak-ro, Daegu 41566, Republic of Korea}

\author[0000-0002-4179-2628]{Hyunjin Shim}
\affiliation{Department of Earth Science Education, Kyungpook National University, 80 Daehak-ro, Daegu 41566, Republic of Korea}

\author[0000-0002-3370-187X]{Eunchong Kim}
\affiliation{International Gemini Observatory/NSF NOIRLab, 670 N. A'ohoku Place, Hilo, Hawai'i, 96720, USA}

\author[0000-0003-3301-759X]{Jeong Hwan Lee}
\affiliation{Research Institute of Basic Sciences, Seoul National University, Seoul 08826, Republic of Korea}
\affiliation{Department of Physics and Astronomy, Seoul National University, 1 Gwanak-ro, Gwanak-gu, Seoul 08826, Republic of Korea}



\begin{abstract}
We present a systematic search for 1696 Green Pea (GP) galaxy candidates in the southern hemisphere selected from the Dark Energy Survey Data Release 2 (DES DR2) 
and provide preliminary results from spectroscopic follow-up observations of 26 targets chosen among them.
Our selection criteria include the colors in $gri$-bands 
and compact morphology in the color composite images. 
The multi-wavelength spectral energy distribution fitting shows that
the selected GP candidates exhibit star formation rates up to several tens $\mathrm{M}_\odot\,\mathrm{yr}^{-1}$.
With the mean stellar mass of $\mathrm{log}\,M_*/\mathrm{M}_\odot=8.6$,
GP candidates are located at roughly 1\,dex above the main sequence of star-forming galaxies at $z\sim0.3$.
Spectroscopic follow-up observations of the GP candidates with Gemini/GMOS are underway. 
All 26 targets are spectroscopically confirmed to be at $z=0.3$-0.41 and have 
[O\textsc{iii}] equivalent width larger than 85\,$\mathrm{\AA}$, classified to be starbursts 
with low to moderate dust attenuation. 
These confirmed GPs show a lower metallicity offset from the mass-metallicity relation of local star-forming galaxies,
indicating that GPs are less chemically evolved systems at their early stage of evolution. 
\end{abstract}



\section{Introduction} \label{sec:intro}

Green Pea galaxies (hereafter GPs), which were first identified 
based on the green color and compact morphology in the SDSS \textit{gri}-composite images \citep{2009MNRAS.399.1191C},
are characterized as
starburst galaxies at $z\sim0.3$ 
with strong [O\textsc{iii}]$\lambda5007$ emission lines 
(corresponding to have large equivalent widths (EWs) up to $\sim$ 1000\r{A}).
Based on the physical properties of GPs studied so far 
$-$ e.g., high star formation rate (SFR) 
for a given stellar mass \citep{2009MNRAS.399.1191C, 2020ApJ...898...68B,2022ApJ...927...57L},
low gas phase metallicity \citep{2010ApJ...715L.128A,2012ApJ...749..185A, 2011ApJ...728..161I},
and small size \citep[$<$ 1kpc; ][]{2017ApJ...838....4Y,2020ApJ...893..134K,2021ApJ...914....2K}
$-$, GPs are considered to be low-redshift analogs of Ly$\alpha$ emitters at $z>2$
with possibilities of Ly$\alpha$ photon escape \citep{2015ApJ...809...19H, 2016ApJ...820..130Y}.
The leakage of the Lyman continuum is also observed in several GPs
through the rest-frame UV observations \citep{2016MNRAS.461.3683I, 2018MNRAS.478.4851I}.
As the similarities between GPs and galaxies in the early Universe are claimed 
by the recent JWST observations
of rest-frame spectra of $z\sim8$ galaxies \citep{2022A&A...665L...4S, 2023ApJ...942L..14R},
GPs are getting attention as tools
to probe the factors and processes related to the cosmic reionization.

GPs have inspired studies of `extreme emission line galaxies (EELGs)' at different redshifts
that have played important roles in galaxy evolution \citep{2016A&A...585A..51D, 2018ApJ...868L..33L,2022MNRAS.513.4451B}.
EELGs are characterized by high excitation nebular spectra
with large EW emission lines.
The observed line ratios in EELGs, 
such as high [O\textsc{iii}]/[O\textsc{ii}], 
are often used as tracers of Ly$\alpha$ and Lyman continuum photon escape
\citep{2014ApJ...791L..19J, 2019ApJ...874...52M, 2019MNRAS.489.2572T}.
EELG selection is also related to a search for extremely metal-poor galaxies \citep{2015MNRAS.454L..41B}, 
especially when there exist possibilities that different metallicity calibrators 
do not agree with each other 
in a low metallicity environment \citep{2010ApJ...715L.128A, 2023MNRAS.518..425C}.

Imaging surveys paralleled with the spectroscopic surveys 
in the northern sky 
\citep[e.g., SDSS and LAMOST;][]{2000AJ....120.1579Y, 2015RAA....15.1095L}
have led to the construction of the large GP sample in the northern hemisphere
\citep[e.g.,][]{2009MNRAS.399.1191C, 2022ApJ...927...57L}. 
However, there have been rare GP samples explored by the southern sky surveys
including the Dark Energy Survey \citep[DES; ][]{2021ApJS..255...20A},
which can be effective for a systematic search for GP candidates.
It is also expected that the upcoming imaging and spectroscopic surveys
such as the Legacy Survey of Space and Time \citep[LSST; ][]{2019ApJ...873..111I}
can contribute to constructing large samples of EELGs (including GPs) in the southern sky.
Another advantage of exploring GPs in the southern sky area is
the existence of surveys and instruments using multiple narrow- and medium-band filters, such as 
J-PLUS \citep{2019A&A...622A.176C, 2022A&A...668A..60L}, 
J-PAS \citep{2021A&A...653A..31B},
and 7-Dimensional telescope \citep[7DT;][]{2021cosp...43E1537I},
which would lead to more efficient selection of EELGs
\citep[e.g.,][]{2022A&A...668A..60L, 2022A&A...665A..95I}. 
Comparison between the narrow- and medium-band filters information-aided selection
and conventional broad-band color selection of strong line emitters
would provide a hint to estimate sample completeness in these rare populations,
complementing spectroscopic selection of EELGs.

In this study, we select GP candidates in the southern hemisphere 
based on the broad-band optical colors %($g-r$ and $r-i$) 
and morphology, 
in advance of the upcoming large and deep spectrophotometric surveys 
using narrow- and/or medium-band filters. 
The physical properties of the selected GP candidates
and the scaling relations between those properties 
(SFR, stellar mass, and gas-phase metallicity) 
are explored through the broad-band spectral energy distribution (SED) 
fitting and follow-up spectroscopic observations.

The paper is organized as follows. 
In Section~\ref{sec:data}, we list our GP candidates selection process 
and the multi-wavelength photometry data of GP candidates compiled for the construction of the SED.
In Section~\ref{sec:method}, 
we describe the SED fitting procedure (Section~\ref{sec:sed})
and our spectroscopic follow-up observations made during the year 2024 
(Section~\ref{sec:spectro}). 
Physical properties (photometric and spectroscopic redshifts, 
[O\textsc{iii}] line EWs, dust attenuation $E(B-V)$, SFR, stellar mass,
and metallicity) and relations between them are presented in Section~\ref{sec:properties}. 
Then we discuss the characteristics of the DES GPs 
through the comparison with the previously studied GP samples 
(constructed mostly in the northern hemisphere)
along with our conclusion (Section~\ref{sec:outro}).
Throughout this paper, 
we use a flat cosmology model with the following parameters:  
$H_0=67.4\,$km\,s$^{-1}$\,Mpc$^{-1}$, 
$\Omega_m=0.315$, and $\Omega_\Lambda=0.685$ \citep{2020A&A...641A...6P}. 
All magnitudes are presented in the AB system.

\section{Sample} \label{sec:data}

\subsection{Green pea candidates selection}

We selected candidates of the GPs
from the second data release of the Dark Energy Survey 
\citep[DES DR2; ][]{2021ApJS..255...20A} covering $\sim5,000$\,deg$^2$,
based on their green colors and compact morphology in the optical images. 
When extracting photometric objects through the SQL query from 
the DESaccess\footnote{https://des.ncsa.illinois.edu/desaccess/}, 
we applied 
constraints for the object flags 
(\texttt{flags\_[\textit{grizY}]} $<$ 2 and \texttt{imaflags\_iso\_[\textit{grizY}]} = 0) 
to exclude objects that are affected by nearby bright stars. 
We also used 
the morphology constraint to remove stellar objects
(using \texttt{wavg\_spread\_model\_[\textit{i}-band]}; \citealt{2018MNRAS.481.5451S}).

\begin{figure}
\epsscale{1.2}
\plotone{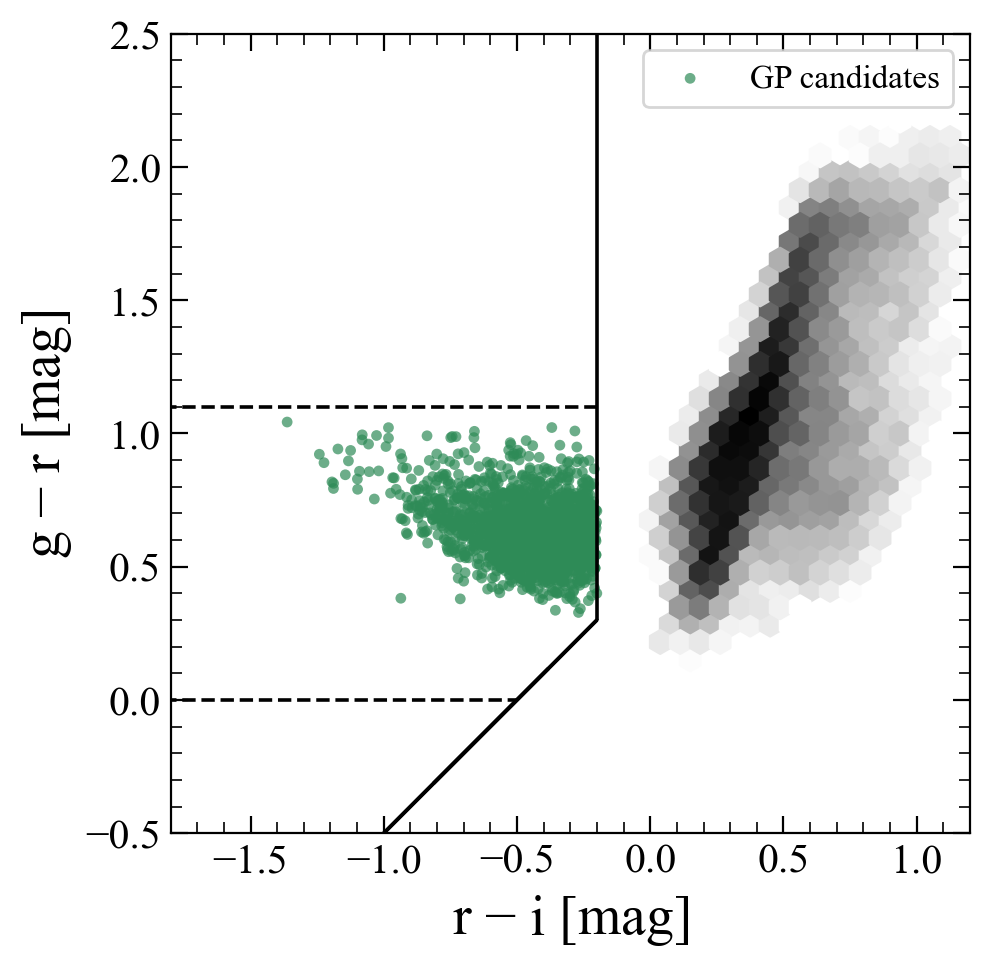}
\caption{\label{fig:colorcut}  
($g-r$) vs. ($r-i$) color-color diagram constructed using DES photometry. 
Black hexagonal binned plots are randomly selected 0.1\% of DES DR2 galaxies within the 
$r$-band magnitude range of $18.0\leq r \leq 22.5$ (same as GPs),
while green circles indicate GP candidates. 
Black solid lines indicate color selection criteria for GPs from \citet{2009MNRAS.399.1191C}.
Black dashed lines are the additional color cut we applied in this work, 
to avoid red galaxies. }
\end{figure}

The extraction was also limited to objects that are observed in all of the five bands of the DES 
(i.e., \texttt{nepoch\_[\textit{grizY}]} $>$ 0),
having Petrosian radius less than or equal to 2\,arcsec, 
considering the compactness of the green pea galaxies \citep{2021ApJ...914....2K}. 
Then we applied the color selection criteria for GPs
as follows \citep[the same as presented in][with the additional constraint for $(g-r)$ to avoid red galaxies]{2009MNRAS.399.1191C}: 
\begin{equation} \label{eqn:selection}
\begin{aligned}
    r - i \leq -0.2, \\
r - z \leq 0.5, \\
g - r \geq r - i + 0.5, \\
0 \leq g - r \leq 1.1. \\
\end{aligned}
\end{equation}

For magnitudes in Equation \ref{eqn:selection}, 
we used MAG\_AUTO from the SExtractor (\texttt{mag\_auto\_[\textit{grizY}]}). 
While the SDSS and DES surveys utilize similar filter sets ($g, r, i,$ and $z$-bands), 
the actual filter responses are slightly different between the two surveys,
causing magnitude offsets between the SDSS and DES bands. 
Therefore, to use the criteria of \citet{2009MNRAS.399.1191C},
we applied a photometric transformation to the DES magnitudes 
using the formula from \citet{2021ApJS..255...20A}
to convert the magnitudes into SDSS magnitudes.
In the case of the GP candidates, the magnitude offsets
(i.e., $|g_\mathrm{SDSS}-g_\mathrm{DES}|$) were mostly smaller than 0.05\,mag. 
In the magnitude range of $18.0\leq r \leq 22.5$, 
the number of GP candidates satisfying the above color selection criteria was 6454.
Figure~\ref{fig:colorcut} describes
color selection criteria for GPs in the ($g-r$) vs. ($r-i$) color-color diagram.

Then we performed a visual inspection of the $gri$ color-composite images 
of the color-selected GP candidates 
to exclude image artifacts (such as satellite trails), 
objects that are not as compact as pea-like galaxies, 
and objects with reddish colors. 
Stars were removed from the GP candidates by searching for matches to 
Gaia Data Release 3 parallax and proper motion measurements \citep{2023A&A...674A...1G}.
The number of GP candidates after the visual inspection and star removal 
is 1696. These 1696 GP candidates comprise our DES GP (candidates) sample studied in this work.

\begin{figure}
\epsscale{1.2}
\plotone{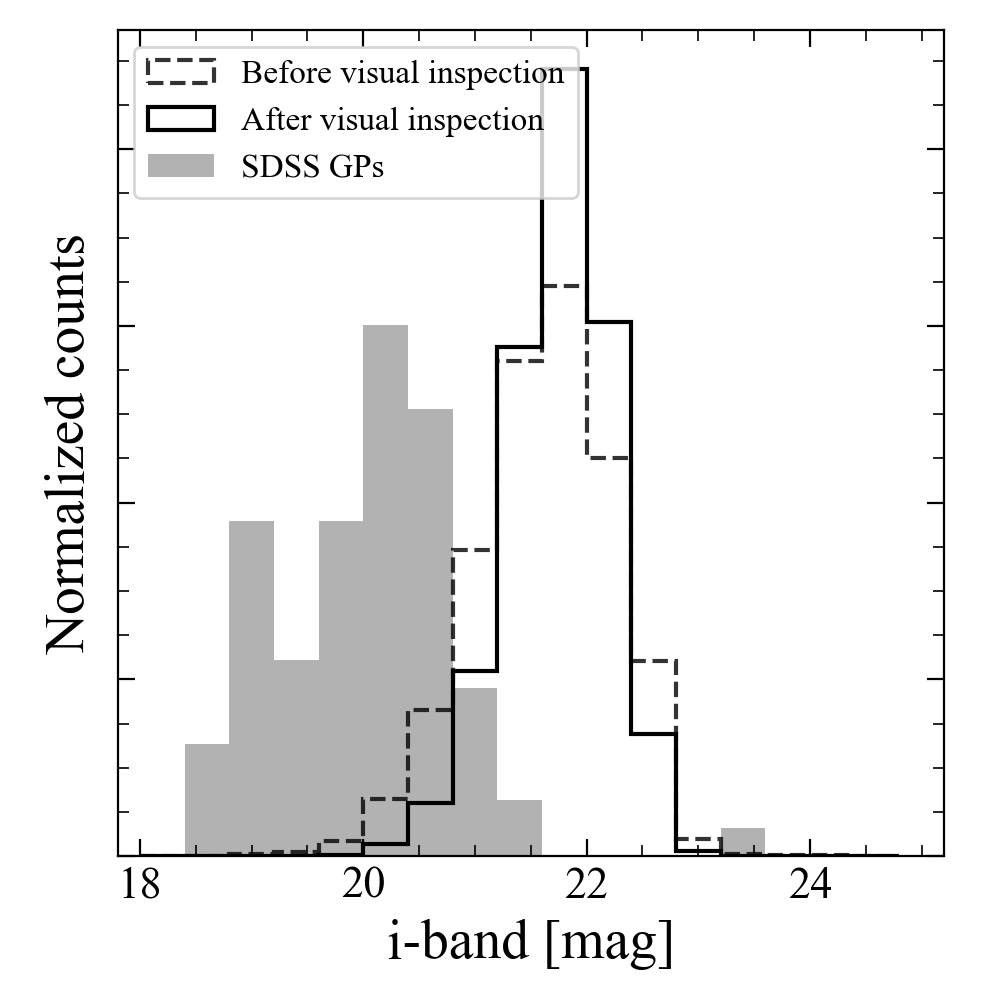}
\caption{\label{fig:iband_hist}
The $i$-band magnitude distributions
of our DES GP candidates (selected in this study) and 80 SDSS GPs \citep{2009MNRAS.399.1191C}.
The dashed and solid line represent the magnitude distributions 
before and after the visual inspection, respectively. 
Gray-filled histogram shows the SDSS GPs.
Note that the total area under the histogram equals 1 in all cases.}
\end{figure}

Figure \ref{fig:iband_hist} shows the $i$-band magnitude distribution of 
the DES GP candidates (before and after the visual inspection) 
compared to that of the SDSS GPs from the early data release \citep[DR7;][]{2009MNRAS.399.1191C}. 
The magnitude distributions
before and after the visual inspection are consistent with each other 
in all optical bands ($g$, $r$, $i$, and $z$-band),
implying that the visual inspection does not result in a sample selection 
biased to the bright objects.
The DES GP candidates we selected are fainter than the SDSS GPs, 
which is natural considering the high fraction of spectroscopically observed sources 
among the SDSS GPs. The mean $i$-band magnitudes of the DES and SDSS GPs 
are 21.75 and 20.20 mag, respectively.  
The $\sim$1.5 mag difference on average suggests that GP candidates selected in the DES 
are fainter (and thus less massive) versions of the conventional SDSS GPs.

\begin{figure}
\epsscale{1.2}
\plotone{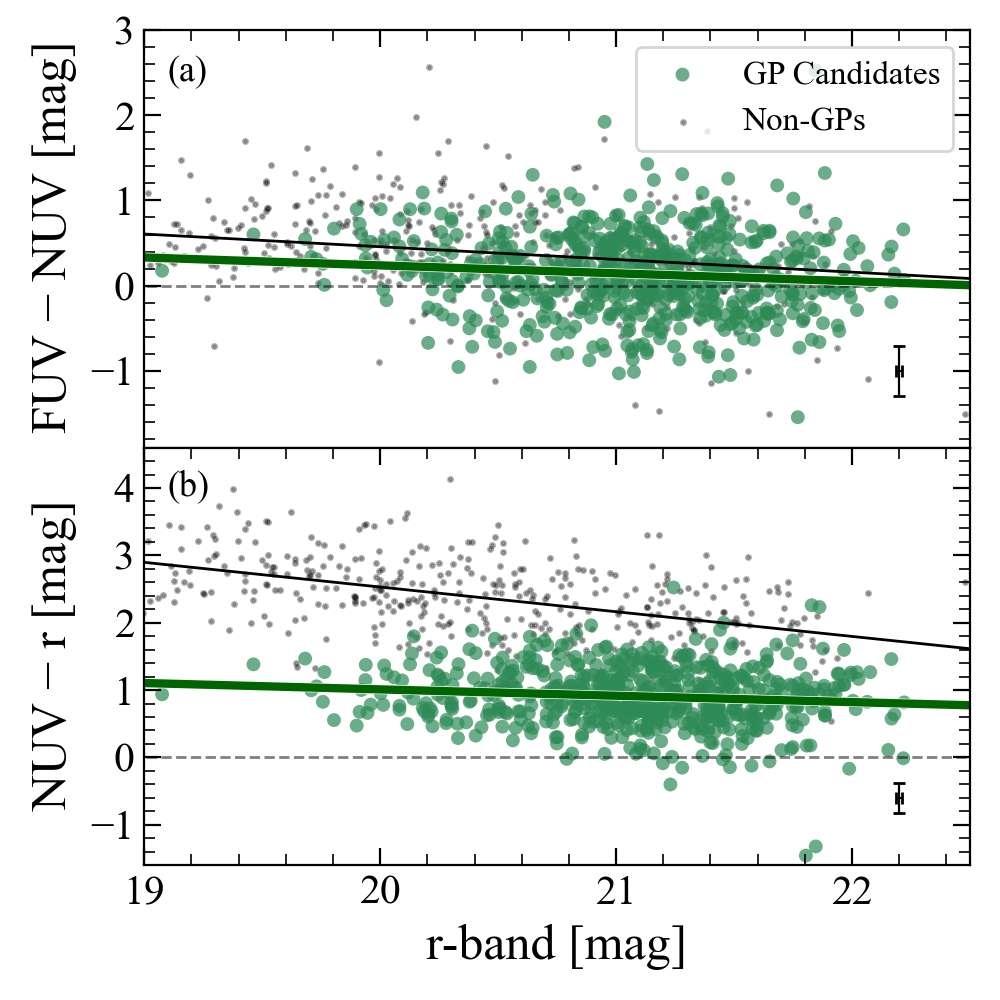}
\caption{\label{fig:uv_colors}
(a) (FUV$-$NUV) color and (b) (NUV$-$$r$) color vs. $r$-band magnitude diagram 
constructed using GALEX and DES photometry. 
Green circles and solid lines represent GPs and their trend lines,
while black points and solid lines indicate Non-GPs and their trend lines, respectively.
Error bars in the bottom right of each panel
represent average errors in colors.}
\end{figure}

\subsection{Multi-wavelength photometry}

To construct SEDs of the GP candidates, 
we compiled multi-wavelength photometry data from the ultraviolet (UV) to near-infrared (NIR)
wavelengths. 
UV magnitudes of the DES GP candidates 
were extracted from the source catalog of the 
GALEX All-sky Imaging Survey Data Release 6 \citep{2014yCat.2335....0B}
using a matching radius of 3\,arcsec. 
The total number of the matched sources is 1176,
while the number of sources with available fluxes at both FUV ($1500$\,\r A) and NUV ($2300$\,\r A) bands is 604. 

To compare the UV photometric properties of the DES GPs with those of other galaxies, 
we selected 50000 non-GP galaxies from the DES with the same $i$-band magnitude distribution 
as the GP candidates, of which 467 galaxies have flux measurements in both FUV and NUV bands. 
Figure~\ref{fig:uv_colors} compares color-magnitude diagrams of the GP candidates
and magnitude-matching non-GPs in the UV and optical wavelengths. 
GP candidates are slightly brighter in the UV wavelengths
(median magnitudes of 21.53 and 22.04\,mag in FUV and NUV bands)
than non-GPs (median magnitudes of 23.18 and 22.64\,mag in FUV and NUV bands, respectively).
In addition, the (FUV$-$NUV) colors of the GP candidates are on average
0.3\,mag bluer than those of the non-GPs at $r=20$\,mag,
with little dependence on the $r$-band magnitudes for the color offset. 
The color difference is even more significant in the case of (NUV$-r$) colors:
non-GP galaxies exhibit the color-magnitude relation 
(i.e., brighter galaxies are redder), however, 
such a trend is relatively weak in the case of GP candidates. 
This implies that GP candidates are bluer galaxies
compared to non-GP galaxies,
because of being 
less dust-attenuated and/or having younger stellar populations.

NIR magnitudes of the GP candidates in 1-4\,$\mu$m were compiled 
to estimate stellar masses of them based on the multi-wavelength SED fitting. 
We used two catalogs:
(1) the VISTA Hemisphere Survey (VHS) Data Release 6 \citep{2021yCat.2367....0M}
for photometry in J, H, and K-bands,
and (2) the CatWISE2020 catalog \citep{2021ApJS..253....8M} 
for photometry in W1 ($3.4\,\mu$m) and W2 ($4.6\,\mu$m) bands. 
Using the matching radius of 1\,arcsec, the number of the identified GP candidates 
in the VHS DR6 is 471, 
while most of them (468) are detected only in the J-band. 
We used a larger matching radius (2\,arcsec) to find counterparts in the CatWISE2020 catalog
since the WISE spatial resolution is relatively poor.
In the CatWISE2020 catalog, we found matches for 389 GP candidates.

\begin{figure}
\epsscale{1.2}
\plotone{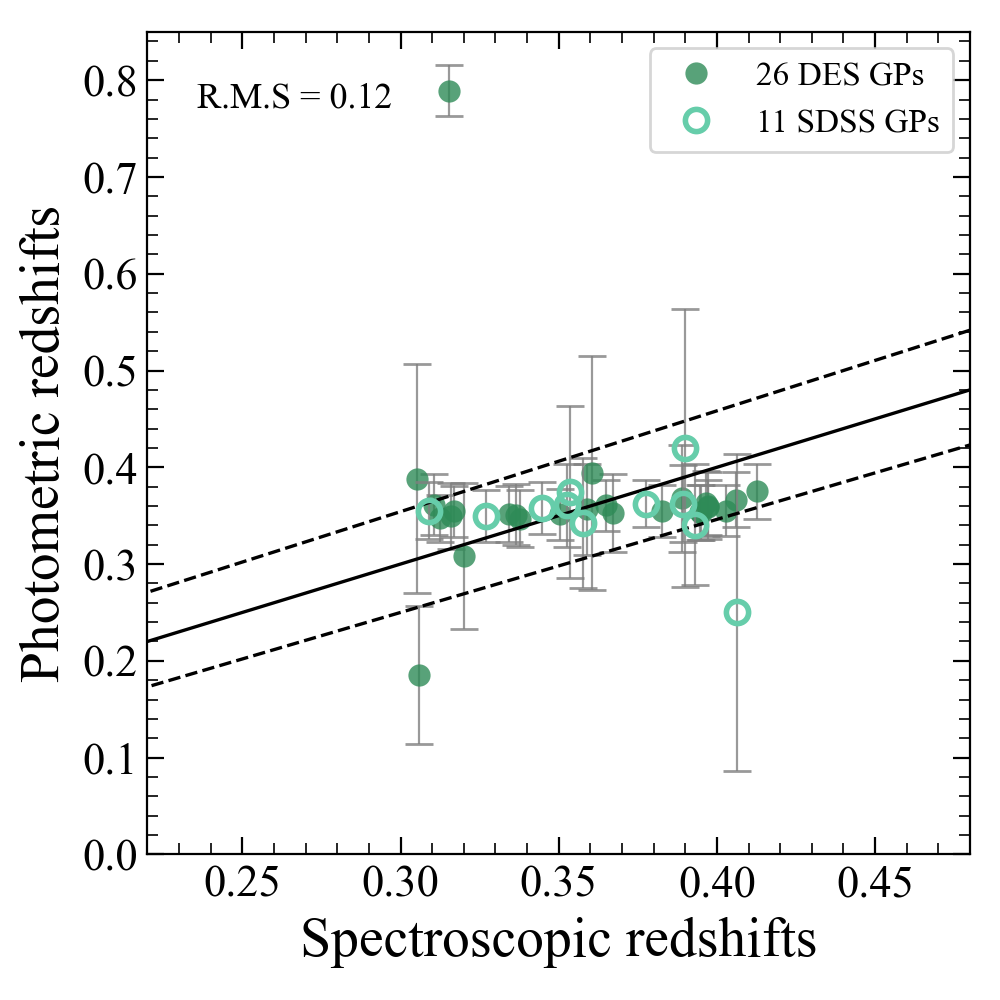}
\caption{\label{fig:specz}
Comparison between spectroscopic redshifts and photometric redshifts derived from \textsc{cigale} SED fitting. 
The solid line shows the $y=x$ equation 
and the dashed lines show $|\Delta z|/(1+z)<0.04$.
Most of the spectroscopically observed GPs show 
$|\Delta z|/(1+z)<0.04$, except for four cases including those with 
$z_\mathrm{phot}<0.26$ or $z_\mathrm{phot}>0.4$.
} 
\end{figure}

\begin{figure}
\epsscale{1.2}
\plotone{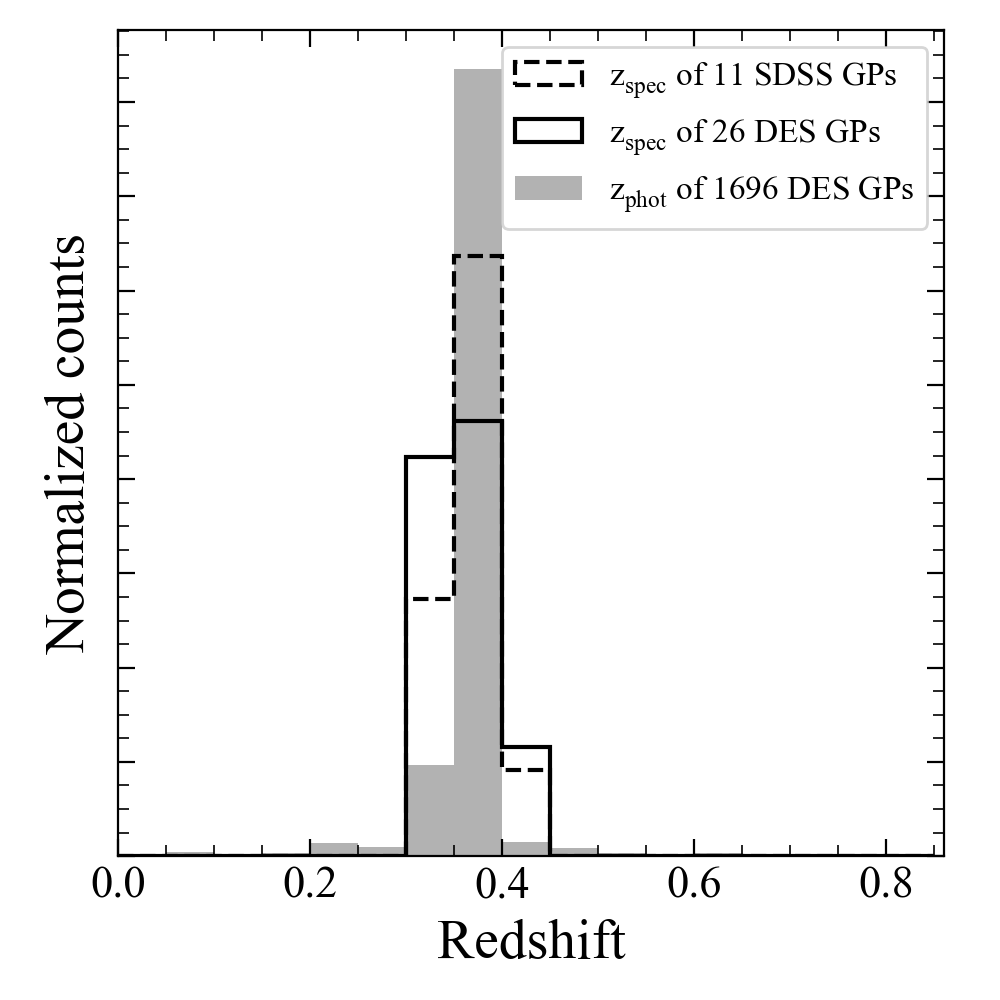}
\caption{\label{fig:photz}
Redshift distributions of SDSS and DES GPs (for both spectroscopic and photometric redshift).
Note that the histograms are normalized so that the total area under the histogram
is equal to 1.}
\end{figure}

\begin{figure*}
\epsscale{1.1}
\plotone{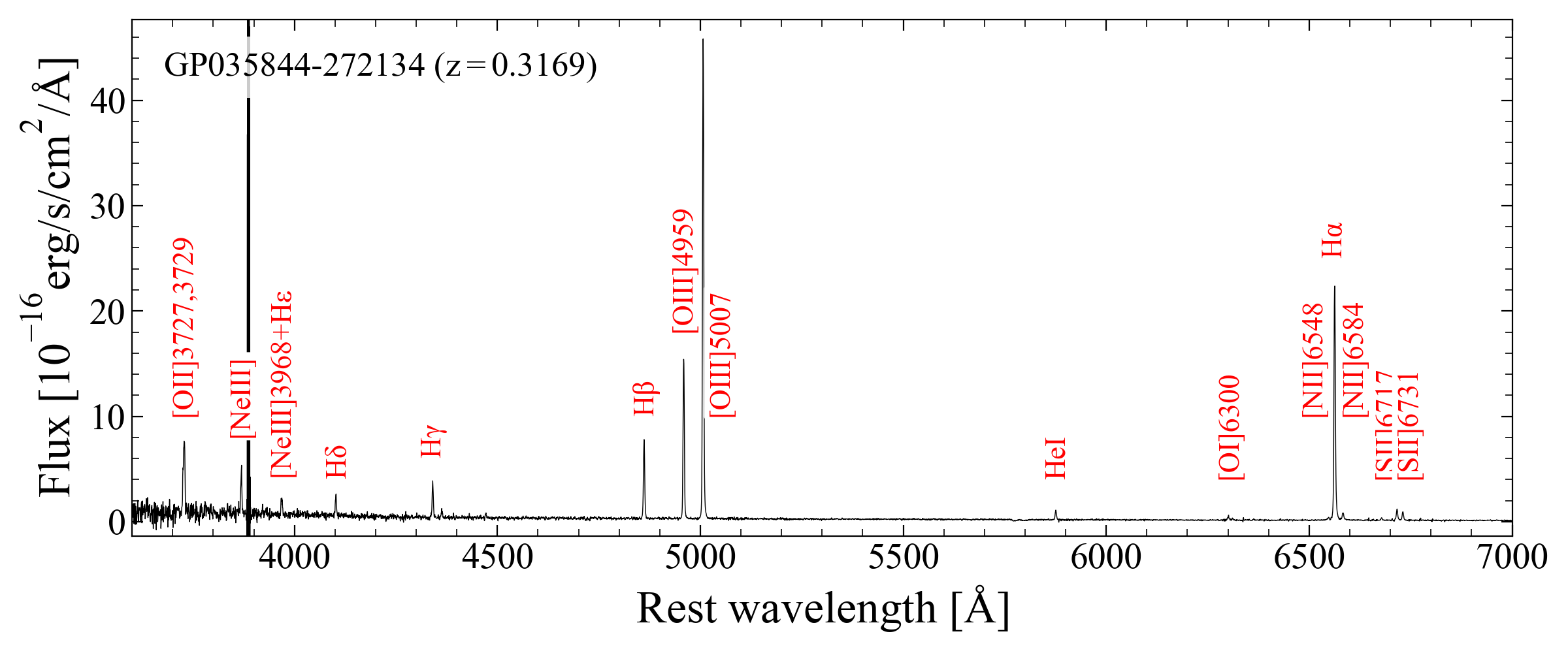}
\caption{\label{fig:spectrum}
The spectrum of the GP035844-272134, which has the strongest line fluxes among 
26 GPs in our Gemini GMOS-S spectroscopic follow-up program, 
is presented as an example in the rest-frame wavelength range of 3600-7000 $\mathrm{\AA}$.}
\end{figure*}

\section{Methods}  \label{sec:method}
\subsection{Multi-wavelength SED fitting} \label{sec:sed}

We estimated the physical properties of the GP candidates,
including SFR, stellar mass, 
E(B$-$V), and equivalent width (EW) of the [O\textsc{iii}] emission line, 
through the multi-wavelength SED fitting with 
the use of \textsc{cigale} \citep{2019A&A...622A.103B}.
We used all of the compiled photometric data points that are available, i.e., 
magnitudes in the FUV, NUV, $g$, $r$, $i$, $z$, $Y$, J, H, K, W1, and W2-bands. 
The number of GP candidates with magnitudes 
in more than 6 out of the 12 photometric bands is 1318,
while 412 GP candidates have fluxes in more than 8 photometric bands. 

To determine the most appropriate parameters and configurations 
for the SED fitting, 
we used 11 objects among our 1696 GP candidates
that have been spectroscopically observed in SDSS DR18.
These 11 GPs have spectroscopic redshifts of $z=0.35$-0.42,
with strong [O\textsc{iii}] emission lines present in their spectra.
By changing the modules and parameter values (ranges and steps) in each module, 
we tried to reproduce the spectroscopic redshifts and [O\textsc{iii}]$\lambda$5007 EWs 
of these objects using the photometric redshift mode of the \textsc{cigale}. 
In the photometric redshift mode, the photometric redshifts were obtained using a Bayesian analysis 
based on the SED fitting, searching for the best-fitting redshift in the range of 
$z=$0.05 to 1.0 at intervals of 0.01. 

The finally selected modules and parameter value settings 
are summarized in Table~\ref{tab:sedfit}. 
We used a delayed star formation history (\texttt{sfhdelayed}),
\texttt{bc03} stellar population synthesis model \citep{2003MNRAS.344.1000B}, 
Chabrier initial mass function \citep{2003PASP..115..763C}, 
nebular emission and dust attenuation. 
Although we used modules for AGN emission (\texttt{fritz2006}) and dust emission (\texttt{dl2014}), 
we fixed the AGN luminosity fraction to be zero
and did not use a range of parameters in the dust emission module, 
because the AGN and dust emission modules can only be constrained by the photometry 
at wavelengths longer than the mid-infrared. 
In the nebular emission module, we used a wide range of ionization parameter $U$,
gas metallicity, and line width to reproduce the spectroscopic EW. 
With these settings, 
we could derive photometric redshifts with the accuracy of $|\Delta z|/(1+z)<0.04$ at $0.3<z<0.45$
(Figures~\ref{fig:specz} and \ref{fig:photz}),
which were confirmed later with the additional 26 %\textbf{25? [change]} 
spectroscopic redshifts obtained from our observations (Section~\ref{sec:spectro}).

\subsection{Follow-up spectroscopic observation} \label{sec:spectro}

To test the efficiency of sampling strong [O\textsc{iii}] emitters in the DES region
based on the GP color selection technique, 
we initiated the spectroscopic follow-up observation program for GP candidates 
with the Gemini Multi-Object Spectrograph (GMOS) on Gemini-South, 
utilizing long-slit observing mode at relatively poor weather conditions (Bands 3 and 4). 
While the observation program is still an ongoing project,
the spectra of 26 GP candidates that have been obtained over the period of 
April to September 2024 are presented in this paper.
We used R400 grating with two central wavelength settings (7000\r{A} and 7500\r{A})
to avoid the line falling in the gaps between the CCD chips.
Our spectrum covers the wavelength range of 3600-10300 \r{A}.
With a 0.5\,arcsec width slit and no binning in spectral direction, 
the spectral dispersion was 0.85\r{A}\,pixel$^{-1}$ resulting in $R\sim1800$. 
The on-source integration time was either 2400\,seconds 
(600\,s$\times4$, consisting of four ABBA dithering for sky background removal) 
or 3000\,seconds (750\,s$\times4$),
which was determined based on the $r$-band magnitude of the target. 
 
The spectra were reduced in a standard manner  
(i.e., bias/flat correction, wavelength calibration, distortion correction,
and 1-D spectra extraction) 
using the DRAGONS\footnote{https://zenodo.org/records/10841622} \citep[Data Reduction for Astronomy from Gemini Observatory North and South;][]{2023RNAAS...7..214L}.
The baseline calibration (using spectrophotometric standard star observation) and
the aperture extraction were done in an interactive mode, 
with caution paid to prevent overfitting in sensitivity function calculation and aperture tracing.

Figure~\ref{fig:spectrum}
shows an example of the reduced GP spectrum obtained through our program.
Similarly to the previously confirmed GPs \citep[e.g.,][]{2009MNRAS.399.1191C},
the spectrum shows numerous nebular emission lines, 
including strong [O\textsc{iii}] and Hydrogen Balmer lines.
The coordinates, $r$-band magnitudes, spectroscopic redshifts along with the 
stellar mass and SFR estimated from the multi-wavelength SED fitting 
(Section~\ref{sec:sed}) are summarized in Table~\ref{tab:properties}.
Spectroscopic redshifts of the observed objects range $z=0.3$-0.41.
It is not straightforward to calculate the completeness 
or the purity of our GP selection strategy 
with limited numbers of spectroscopically followed targets.
However, the fact that all targets lie at $z\sim0.3$ 
and show strong [O\textsc{iii}] indicates 
that our selection is effective to search for GPs at $z\sim0.3$
and is relatively free from contaminations of emission-line galaxies at other redshifts.

\begin{figure}
\epsscale{1.2}
\plotone{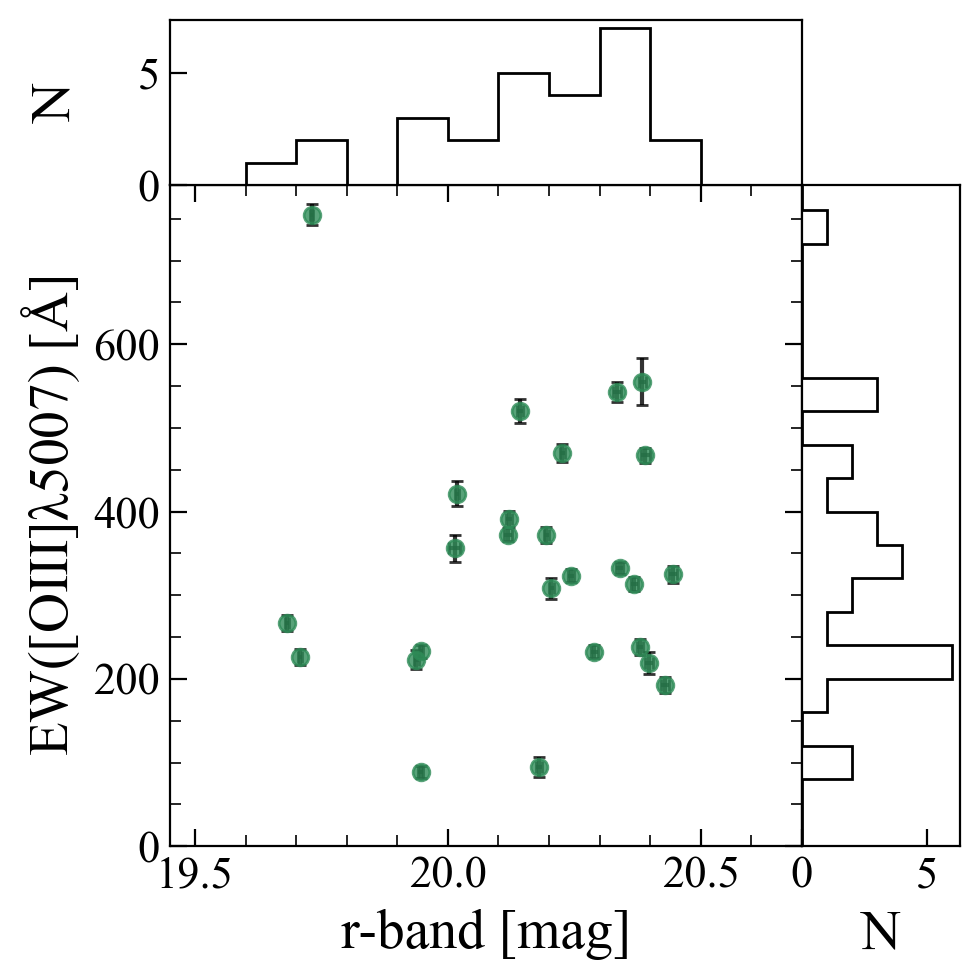}
\caption{\label{fig:ew_r}
[O\textsc{iii}]$\lambda$5007 equivalent width vs. 
$r$-band magnitude for 26 GPs that are spectroscopically observed. 
}
\end{figure}

\begin{figure*}
\epsscale{1.1}
\plotone{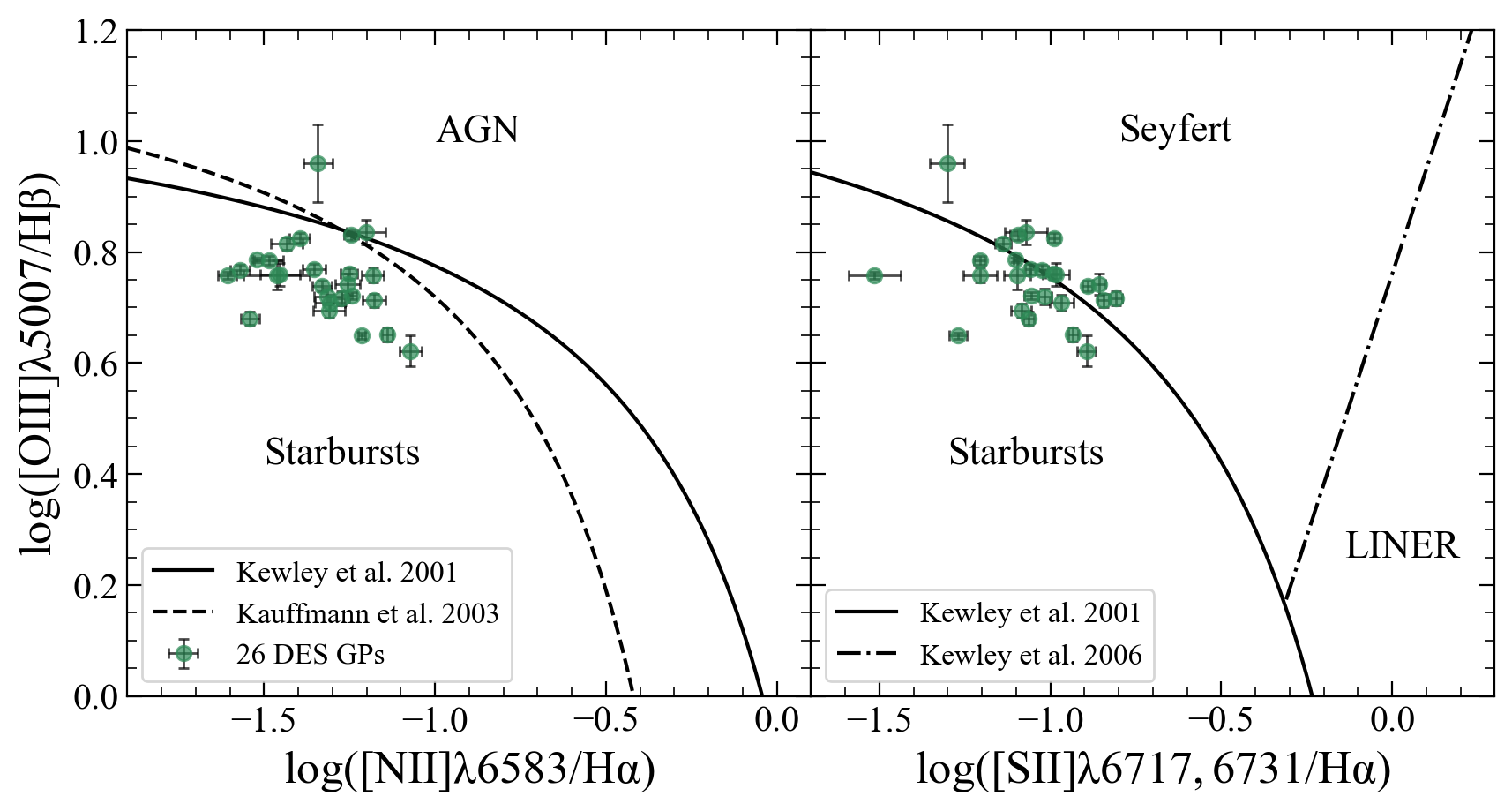}
\caption{\label{fig:bpt}
BPT diagrams \citep{1981PASP...93....5B} to classify galaxies into starbursts and AGN (left), 
as well as into starbursts, Seyferts, and LINER (right). 
Solid and dashed lines are dividers for starbursts and AGN from 
\citet{2001ApJ...556..121K} and \citet{2003MNRAS.346.1055K}, respectively.
Dash-dot line is a suggested criterion to differentiate between 
Seyfert and LINER \citep{2006MNRAS.372..961K}.}
\end{figure*}

\begin{figure}
\epsscale{1.2}
\plotone{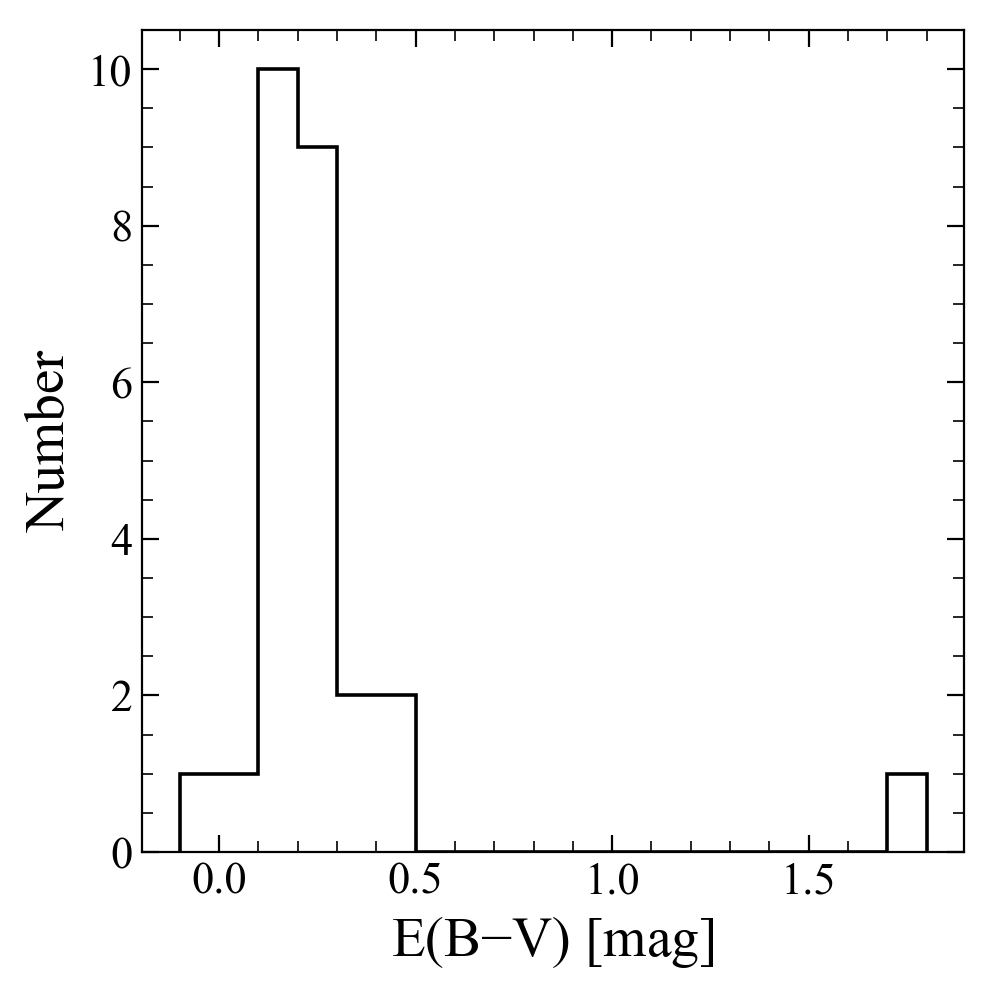}
\caption{\label{fig:extinction}
Color excess $E(B-V)_\mathrm{gas}$ of the spectroscopically observed 26 GPs 
derived from the observed Balmer line ratios (H$\alpha$/H$\beta$).
The \citet{2000ApJ...533..682C} dust attenuation curve was used.       
}
\end{figure}

\section{Results} \label{sec:properties}

\subsection{Redshift distribution}
 
As is discussed in Section~\ref{sec:sed}, 
the photometric redshift accuracy of our SED fitting is $|\Delta z|/(1+z)<0.04$.
In addition to the comparison between spectroscopic and photometric redshifts (Figure~\ref{fig:specz}), 
Figure~\ref{fig:photz} shows the photometric redshift distribution of the 
1696 GP candidates 
as well as spectroscopic redshift distribution of the 
spectroscopically observed GPs. 
Although the spectroscopic target selection is inevitably biased to
the bright targets, the redshift distributions of the spectroscopic subsample
and the parent photometric sample are consistent with each other.
The mean value of the photometric redshift is 
$\langle z \rangle=0.36$, with a standard deviation of 0.05. 

The photometric redshift range of GP candidates selected in the DES 
is slightly higher than that of GPs selected in the SDSS 
\citep[$0.112\leq z \leq 0.360$, ][]{2009MNRAS.399.1191C}.
The difference appears to reflect the difference in filter systems. 
The central wavelengths of the DES $gri$ filters 
are slightly longer (4808, 6417, and 7814\,\AA) 
than that of the SDSS $gri$ filters (4702, 6175, and 7489\,\AA). 
The green color of the GPs in the $gri$ color-composite images
require strong [O\textsc{iii}] emission line that is redshifted into the $r$-band wavelength range, 
which naturally explains GPs selected in the DES images having slightly higher redshifts.

\subsection{[O\textsc{iii}]$\lambda$5007 EWs}

For 26 GP targets of the GMOS follow-up spectroscopic observations,
we measured line flux and EW of noticeable emission lines 
(with the line $S/N>3$) using the 
\textsc{jdaviz/Specviz} tool \citep{2022zndo...5513927D} in the interactive mode
(Tables~\ref{tab:lineflux} and \ref{tab:ew}).
As it is expected from the green color in the $gri$ images, GPs are considered to have 
strong [O\textsc{iii}] emission lines, with EW ranging up to a few hundreds $\mathrm{\AA}$ 
(Table~\ref{tab:ew}). Figure~\ref{fig:ew_r} shows the distribution of the 
[O\textsc{iii}]$\lambda$5007 EW as a function of the $r$-band magnitude,
which does not show any correlation between the two. 
This implies that (although the number of galaxies used here is small) the $r$-band magnitude cut for strong line emitter selection would not place bias in sample selection in terms of the line EW.
This is consistent with the previous studies
suggesting that the fractions of the EELGs do not vary significantly with UV luminosity
among UV-selected star-forming galaxies \citep{2022MNRAS.513.4451B},
considering that large EW corresponds to high ongoing SFR.

The mean value of the observed [O\textsc{iii}]$\lambda$5007 EW is 340$\mathrm{\AA}$.
Although there are numerous different criteria for EW to define line emitters at different redshifts
\citep[e.g.,][]{2015A&A...578A.105A, 2020ApJ...898...45T, 2022A&A...665A..95I, 2024MNRAS.535.1796B},
the mean value of the observed [O\textsc{iii}]$\lambda$5007 EW, 340$\mathrm{\AA}$, 
for DES spectroscopically observed GPs, is comparable to that of 
most strong line emitters.

\subsection{AGN contamination}
Figure~\ref{fig:bpt} shows the BPT diagrams \citep[][]{1981PASP...93....5B} 
to classify the spectral types of GPs into either starburst-dominated or AGN-dominated systems.
Most (23 out of 26) GP candidates are classified as starbursts 
while only two objects (GP211619-463914, GP232739-454554) are classified as AGN 
and one object (GP204359-540359) is lying on the boundary. 
Despite most GPs being classified as starbursts,
the possibility of them being close to Seyferts
cannot be ruled out (right panel of Figure~\ref{fig:bpt}).
However, the possible Seyfert-like objects do not show broad hydrogen recombination lines 
in the reduced spectra.

\subsection{Dust attenuation}
We used the Balmer decrement (observed line flux ratios between H$\alpha$ and H$\beta$; Table~\ref{tab:lineflux}) 
to estimate dust attenuation of the GPs.
Considering that our GPs are mostly star-forming galaxies (Figure~\ref{fig:bpt}),
we used the extinction curve for starburst galaxies \citep{2000ApJ...533..682C}. 
The color excess $E(B-V)$ was calculated using the following formula:

\begin{equation} \label{eqn:extinction}
\begin{aligned}
    E(B-V)_\mathrm{gas} = \frac{2.5\times\mathrm{log_{10}}[(f_{\mathrm{H}\alpha}/f_{\mathrm{H}\beta})/2.86]}{k(\mathrm{H}\beta)-k(\mathrm{H}\alpha)},
\end{aligned}
\end{equation}

\noindent where the $k$(H$\alpha$) and $k$(H$\beta$) values were taken from \citet{2001PASP..113.1449C},
and the intrinsic $(f_{\mathrm{H}\alpha}/f_{\mathrm{H}\beta})$ was considered to be 2.86 
assuming the case B recombination with $T=10^{4}$\,K and $n_e=10^{4}$\,cm$^{-3}$.
Figure~\ref{fig:extinction} shows the $E(B-V)_\mathrm{gas}$ distribution 
of the spectroscopically observed GPs. 
The median value is $E(B-V)_\mathrm{gas}=0.21$, while most GPs showing color excess less than 0.5\,mag
except one object (GP232739-454554) with very low H$\beta$ line flux.
The spectrum of the GP232739-454554 shows lower S/N in the shorter wavelength than in the longer wavelength.
Therefore, the fact that the S/N of the H$\beta$ line being low is thought to be the reason of high attenuation in this object. 
Similarly to previous studies on the attenuation in GPs
\citep[e.g.,][]{2009MNRAS.399.1191C, 2022ApJ...927...57L}, 
our GPs are relatively blue galaxies (which is consistent with their UV to optical colors in Figure~\ref{fig:uv_colors}) 
with little dust attenuation.

\subsection{SFR vs. stellar mass}

\begin{figure}
\epsscale{1.2}
\plotone{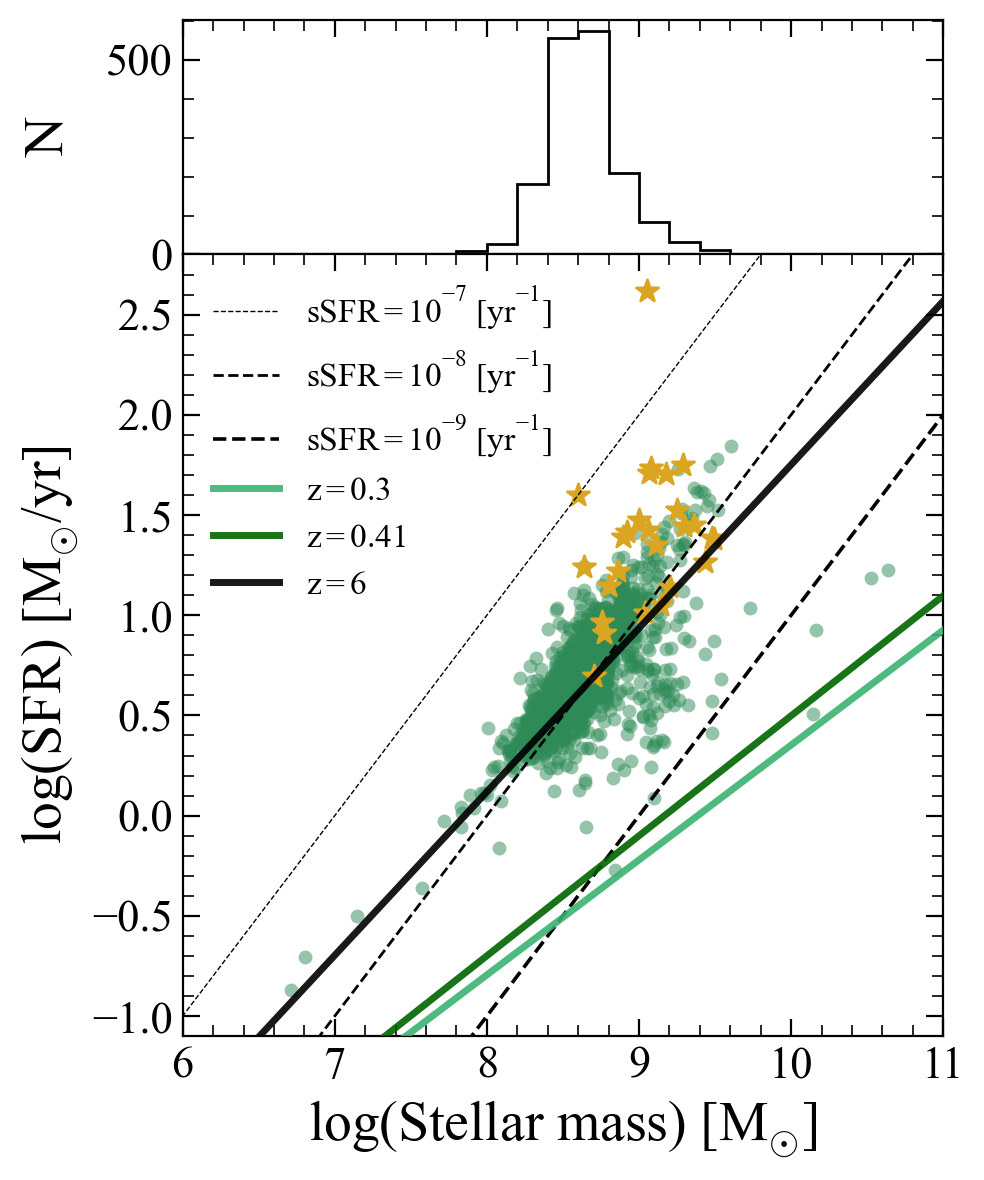}
\caption{\label{fig:sfr_mass} 
The SFR vs. stellar mass for the 1696 GPs estimated by \textsc{cigale}
with the distribution of stellar mass.
The dashed lines represent specific star formation rates (sSFR) of
$10^{-7}, 10^{-8}$ and $10^ {-9} \mathrm{yr^{-1}}$, respectively.
Most of the GPs are located along the sSFR line of $10^{-8} \mathrm{yr^{-1}}$,
with a few GPs positioned along $10^{-9} \mathrm{yr^{-1}}$ line.
Solid lines represent the star-forming main sequence (SFMS) relations with $z=0.3$, 0.41 and 6 
using the redshift evolution of star-forming main sequence 
suggested by \citet{2014ApJS..214...15S}.
Filled stars represent 26 spectroscopically observed GPs,
for which SFR is calculated based on the extinction-corrected H$\alpha$ line flux.
}
\end{figure}

The distributions of stellar mass and SFR of the GP candidates
are presented in Figure~\ref{fig:sfr_mass}. 
The stellar mass of the most GP candidates ranges 
10$^{8}$-10$^{9}$~\(\textup{M}_\odot\),
with a mean value of $\langle\mathrm{M}_\mathrm{star}\rangle=10^{8.6}$~\(\textup{M}_\odot\).
This places the DES GP candidates in the lowest regime of the stellar mass
(10$^{8.5}$-10$^{10}$~\(\textup{M}_\odot\))
among the early SDSS GPs \citep{2009MNRAS.399.1191C},
which is natural considering that DES GPs are $\sim1.5$\,mag fainter 
than that of the SDSS GPs (Figure~\ref{fig:iband_hist}). 

GP candidates exhibit SFR ranging a few to several tens \(\textup{M}_\odot\)$\rm ~yr^{-1}$,
therefore the typical specific SFR (sSFR) value, i.e., SFR divided by the stellar mass, is $\sim10^{-8}$\,yr$^{-1}$. 
Compared to the main sequence of star-forming galaxies at $z=0.3$-0.41 \citep{2014ApJS..214...15S}, 
GP candidates show an order of magnitude higher SFR 
at the same stellar mass.
The sSFR of the GP candidates are comparable to that of $z=6$ star-forming main sequence galaxies,
indicating that the GP selection may lead to a selection of local analogs of high-redshift star-forming galaxies
that are responsible for cosmic reionization.

\begin{figure}
\epsscale{1.2}
\plotone{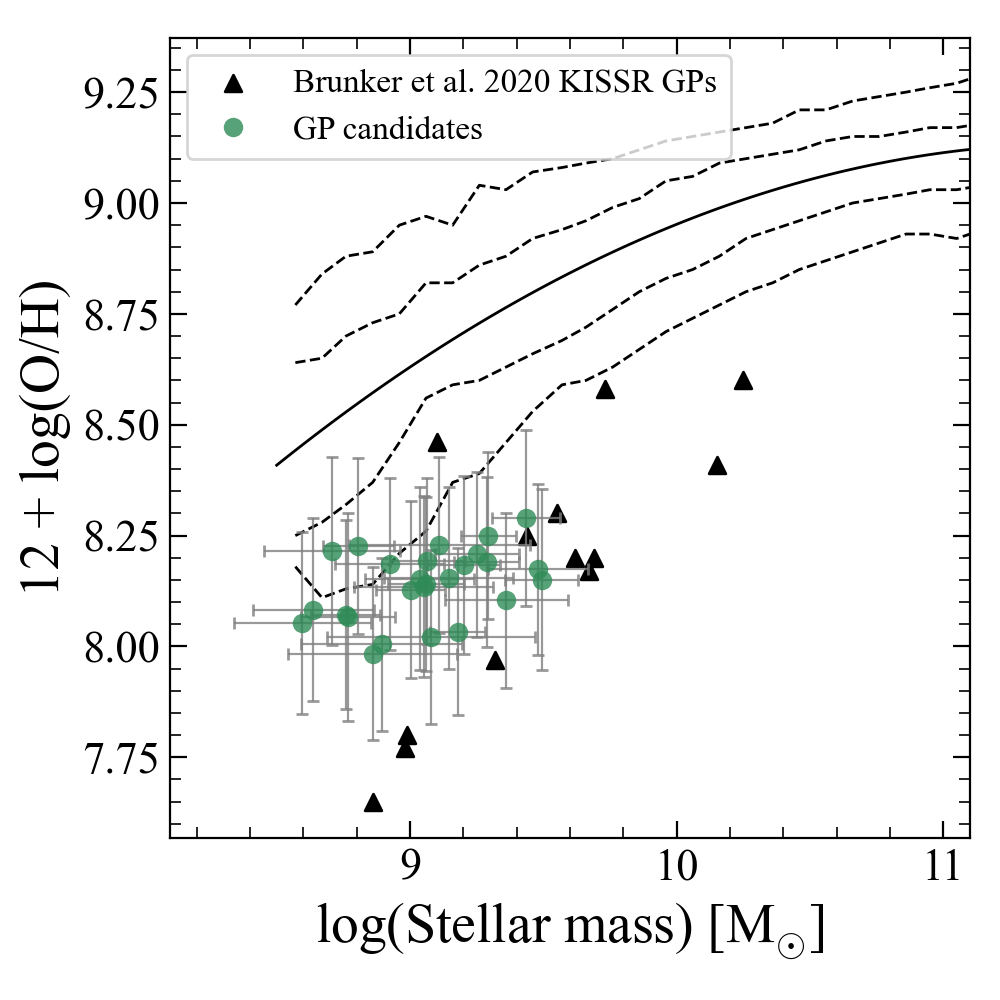}
\caption{\label{fig:mass_metallicity}
Gas-phase metallicity ($12+\mathrm{log(O/H)}$) vs. stellar mass of 26 GPs (green circles). 
Overplotted solid line is the mass-metallicity relation of star-forming galaxies at $z\sim0.1$,
while the dashed lines represent 68\% and 95\% of the mass-metallicity distribution
\citep{2004ApJ...613..898T}.
Filled triangles are GPs from the KPNO International Spectroscopic Survey 
\citep{2020ApJ...898...68B}.
}
\end{figure}

\subsection{Mass-metallicity relation} \label{sec:metallicity}

For 26 spectroscopically observed GPs, we estimated the gas-phase metallicity 
$12+\mathrm{log(O/H)}$ based on the empirical 
$N2$ ($\equiv$ log([N\textsc{ii}]$\lambda$6584/H$\alpha$)) method
\citep[][]{2004MNRAS.348L..59P} that uses strong emission lines:
\begin{equation} \label{eqn:metallicity}
\begin{aligned}
    12+\mathrm{log(O/H)} = 8.90 + (0.57 \times N2).
\end{aligned}
\end{equation}

The values are listed in Table~\ref{tab:properties} along with the 
stellar masses and SFR estimated from the SED fitting. 
GPs have oxygen abundances $12+\mathrm{log(O/H)}<8.3$, 
with the mean value of $\langle 12+\mathrm{log(O/H)} \rangle=8.14$
(i.e., $\sim30$\% the solar value; \citealt{2001ApJ...556L..63A}).
In the mass-metallicity diagram (Figure~\ref{fig:mass_metallicity}), 
GPs lie under the 95\,\% confidence interval
of the local star-forming galaxies \citep[][]{2004ApJ...613..898T}.
At the same time, GPs follow a similar mass-metallicity trend at low metallicities 
that more massive systems follow at higher metallicities.
Such a trend, as well as gas phase metallicity range, is consistent 
with the previous studies on the [O\textsc{iii}]-selected GPs \cite[e.g.,][]{2020ApJ...898...68B}. 
This supports the idea that 
GP selection is an efficient strategy to sample low metallicity galaxies, 
as the strong line emitters tend to be 
chemically less evolved systems
undergoing their early stage of star formation.

\section{Discussion \& Conclusion} \label{sec:outro}

We selected GP candidates from the DES DR2 that covers $\sim5,000$\,deg$^2$ of the southern sky,
and investigated their properties using the multi-wavelength SED fitting. 
In total, 1696 GP candidates were selected by applying broad-band color criteria 
\citep[][]{2009MNRAS.399.1191C} and through visual inspection on the $gri$ color composite images. 
Using the multi-wavelength photometry covering from UV to NIR, 
GP candidates were found to be low-mass ($10^{6.71}$-$10^{10.63}$\,M$_\odot$), 
highly star-forming ($\mathrm{log(SFR)}=-0.87$-1.85) galaxies 
at the mean photometric redshift of $\langle z_\mathrm{phot}\rangle=0.36$. 
Early results from our spectroscopic follow-up program for DES GPs 
suggest that all 26 GP targets do show strong [O\textsc{iii}] emission lines 
with EW larger than $\sim85$\,$\mbox{\AA}$, 
related to a low metallicity (less than 30\,\% the solar metallicity) 
and insignificant attenuation in these systems.

Since our GP candidates sample is one of the starting points to construct 
strong line emitters in the southern hemisphere where numbers of large imaging/spectroscopic surveys 
are planned, we list major characteristics of our GP candidates 
that are common with or 
different from the previously constructed GP samples 
(e.g., SDSS DR7, \citealt{2009MNRAS.399.1191C}; SDSS DR13, \citealt{2019ApJ...872..145J, 2023ApJ...945..157H};
SDSS DR16, \citealt{2023AJ....166..133D}; KISS, \citealt{2020ApJ...898...68B}; LAMOST DR9, \citealt{2022A&A...668A..60L}).
\\

1. DES GPs are expected to lie at slightly higher redshifts than SDSS GPs, 
since the central wavelengths of the DES $g$, $r$, and $i$-bands are
longer than that of the corresponding SDSS bands. 

2. DES GPs are on average less massive ($10^{8.6}$\,M$_\odot$) than SDSS GPs ($10^{9.5}$\,M$_\odot$),
which is natural that DES GPs are $\sim1.5$\,mag fainter. 
While the DES GPs occupy the lowest mass regime among the GP population, 
the sSFR remains to be relatively consistent over the two orders of the stellar mass range. 

3. [O\textsc{iii}] EWs of the spectroscopically observed DES GPs are comparable to 
that of SDSS GPs. The lack of the extremely large EW (reaching up to $\sim1000$\,$\mathrm{\AA}$) galaxies 
might be due to the small size of the targets.
Similarly in the case of the SDSS GPs, little correlation exists
between the $r$-band magnitudes and [O\textsc{iii}] EW. 

4. Judging from the spectroscopically observed DES GPs, 
starburst-dominated systems dominate the GP population rather than AGN-dominated systems.
The AGN fraction is uncertain (2-4 out of 26), 
however the value partly overlaps with that of the search for MIR AGN among SDSS GPs.

5. Spectroscopically observed DES GPs show mass-metallicity relation over $\sim1$ order of 
stellar mass range that more massive GPs show higher gas-phase metallicity. 
This mass-metallicity relation is about 0.5\,dex lower than from the mass-metallicity relation
of local star-forming galaxies, and is consistent with the SDSS and KISSR GPs. \\

The spatial number density of the DES GPs is 0.34\,deg$^{-2}$,
considering the number of GPs is 1696 from the 5000\,deg$^{2}$ covered in the DES DR2. 
This is smaller than that of the early SDSS GPs \citep[about 2\,deg$^{-2}$;][]{2009MNRAS.399.1191C}
or emission line searches in the 
wider redshift range \citep[e.g., $0.11<z<0.93$, 30-100\,deg$^{-2}$;][]{2015A&A...578A.105A, 2022A&A...665A..95I},
yet is comparable or larger than 
that of the local ($z<0.06$) extreme line emitters 
\citep[e.g., 0.0003-0.23\,deg$^{-2}$][]{2017ApJ...847...38Y, 2022A&A...668A..60L}. 
While it is difficult to quantify the sample completeness considering that
broad-band magnitude limits as well as 
EW limits that can be probed by different surveys are different, 
the $r$, $i$, and $z$-band magnitude distributions of the DES GPs suggests that 
the number of the GPs in a specific magnitude bin increases as the magnitude gets fainter.
Considering these, the number of GPs (as well as extreme line emitters) would increase 
through the deep imaging surveys planned in the southern hemisphere \citep[e.g., LSST;][]{2019ApJ...873..111I}
that also cover wide, unexplored area. 
Our DES GP sample, although photometrically selected yet expected to have high purity 
judged by the spectroscopic follow up, 
could be used as a reference data set to complement the future selection strategy for 
strong line emitters in the southern sky.

\begin{deluxetable*}{ll}
    \tablecaption{\label{tab:sedfit}Parameters for \textsc{cigale} SED fitting}
    \tablehead{ \colhead{Modules and input parameters} & \colhead{Range} }
    \startdata
    Star formation history: \texttt{sfhdelayed}   &     \\
    e-folding time of the main stellar population model [Myr]  &  2000  \\
    Age of the main stellar population [Myr]                   &  50, 100, 200, 500,700, 1000, 1500, 2000, 4000, 6000, 8000, 12000  \\
    e-folding time of the late starburst population model [Myr]&  50  \\
    Age of the late burst [Myr]                                &  20  \\
    Mass fraction of the late burst population                 &  0.0, 0.15  \\
    \hline 
    Stellar population: \texttt{bc03}   &    \\
    Initial mass function        & Chabrier   \\
    Metallicity                  & 0.004, 0.008, 0.02 \\
    \hline
    Nebular emission: \texttt{nebular}   &    \\
    Ionization parameter         & -4.0, -3.0, -2.8, -2.6, -2.4, -2.2, -1.6, -1.2, -1.0 \\
    Gas metallicity              & 0.004, 0.008, 0.014, 0.019 \\
    Electron density             & 10, 100, 1000  \\
    Fraction of Lyman continuum photons escaping the galaxy  & 0.0  \\
    Fraction of Lyman continuum photons absorbed by dust  & 0.0  \\
    Line width [km/s]            & 25, 50, 75, 100  \\
    Include nebular emission     & True  \\
    \hline
    Dust attenuation: \texttt{dustatt\_modified\_starburst}   &    \\
    E(B-V) lines                 & 0.0, 0.05, 0.1, 0.2, 0.3, 0.4, 0.5 \\
    \enddata
\end{deluxetable*}

\begin{deluxetable*}{lrrcccccc}
    \tabletypesize{\scriptsize}
    \tablecaption{\label{tab:properties} Properties of the 26 spectroscopically observed DES-GPs}
    \tablehead{
    \colhead{GP ID} & \colhead{R.A.} & \colhead{Decl.} & \colhead{$m_r$} &
    \colhead{$z_\mathrm{spec}$} & \colhead{$\log\,(\textup{M}_\star)$} &
    \colhead{$\log\,$(SFR)$_\mathrm{SED}$} & \colhead{$\log\,$(SFR)$_\mathrm{H\alpha}$} &
    \colhead{$12+\mathrm{log\,(O/H)}$} \\
    \colhead{} & \colhead{[deg]} & \colhead{[deg]} & \colhead{[mag]} &
    \colhead{} & \colhead{$[\textup{M}_\odot]$} &
    \colhead{$[\textup{M}_\odot\,\mathrm{yr^{-1}}]$} & \colhead{$[\textup{M}_\odot\,\mathrm{yr^{-1}}]$} & \colhead{} \\
    \colhead{(1)} & \colhead{(2)} & \colhead{(3)} & \colhead{(4)} &
    \colhead{(5)} & \colhead{(6)} &
    \colhead{(7)} & \colhead{(8)} & \colhead{(9)}
    }
    \startdata
    GP000207$-$424024 & 0.52965 & -42.67360 & 20.02 & 0.3972 & 9.25 & 1.35 & 1.52 & 8.21 \\
    GP000957+024146   & 2.48999 & 2.69620 & 19.94 & 0.3379 & 9.29 & 0.89 & 1.44 & 8.25 \\
    GP001034$-$484901 & 2.64247 & -48.81700 & 19.68 & 0.3059 & 9.08 & 0.24 & 1.73 & 8.02 \\
    GP005423+033009   & 13.59997 & 3.50259 & 19.95 & 0.3590 & 9.44 & 0.81 & 1.27 & 8.29 \\
    GP012256$-$154025 & 20.73453 & -15.67363 & 20.24 & 0.3160 & 9.06 & 0.72 & 1.43 & 8.14 \\
    GP022429$-$105748 & 36.12384 & -10.96339 & 20.33 & 0.3154 & 9.36 & 1.63 & 1.45 & 8.10 \\
    GP024352+035504   & 40.96804 & 3.91804 & 20.37 & 0.3890 & 8.92 & 0.79 & 1.42 & 8.19 \\
    GP031823$-$412811 & 49.59743 & -41.46982 & 20.12 & 0.3972 & 9.06 & 0.83 & 1.71 & 8.19 \\
    GP033528$-$562307 & 53.86939 & -56.38553 & 20.38 & 0.3825 & 8.76 & 0.83 & 0.97 & 8.07 \\
    GP035844$-$272134 & 59.68538 & -27.35972 & 19.73 & 0.3169 & 9.18 & 0.83 & 1.70 & 8.03 \\
    GP042045$-$600735 & 65.18795 & -60.12648 & 20.19 & 0.3651 & 9.00 & 0.73 & 1.48 & 8.13 \\
    GP043224$-$383800 & 68.10077 & -38.63352 & 19.95 & 0.3200 & 9.48 & 0.57 & 1.39 & 8.17 \\
    GP051716$-$265543 & 79.31926 & -26.92878 & 20.23 & 0.3505 & 8.60 & 0.89 & 1.60 & 8.05 \\
    GP052936$-$362507 & 82.40206 & -36.41871 & 20.12 & 0.3052 & 8.89 & 1.03 & 1.39 & 8.00 \\
    GP054214$-$354139 & 85.56175 & -35.69434 & 20.29 & 0.3605 & 9.11 & 1.16 & 1.35 & 8.23 \\
    GP204045$-$413555 & 310.19069 & -41.59872 & 20.39 & 0.4030 & 8.86 & 1.29 & 1.22 & 7.98 \\
    GP204359$-$540359 & 310.99793 & -54.06650 & 20.14 & 0.4126 & 9.29 & 1.35 & 1.75 & 8.19 \\
    GP210903$-$444951 & 317.26582 & -44.83095 & 20.43 & 0.3105 & 8.77 & 0.93 & 0.91 & 8.07 \\
    GP211040$-$504353 & 317.66717 & -50.73162 & 20.34 & 0.3126 & 8.64 & 1.24 & 1.24 & 8.08 \\
    GP211619$-$463914 & 319.08056 & -46.65399 & 20.38 & 0.3951 & 8.71 & 1.03 & 0.70 & 8.22 \\
    GP212733$-$424114 & 321.88912 & -42.68739 & 20.44 & 0.4060 & 9.15 & 1.36 & 1.05 & 8.15 \\
    GP213914$-$495028 & 324.81073 & -49.84135 & 20.20 & 0.3947 & 8.81 & 0.87 & 1.15 & 8.23 \\
    GP215810$-$615644 & 329.54441 & -61.94577 & 20.40 & 0.3364 & 9.20 & 0.66 & 1.14 & 8.18 \\
    GP232739$-$454554 & 351.91373 & -45.76512 & 20.18 & 0.3965 & 9.05 & 1.36 & 2.62 & 8.13 \\
    GP234638$-$011839 & 356.65857 & -1.31104 & 19.71 & 0.3671 & 9.49 & 0.87 & 1.38 & 8.15 \\
    GP234647+022744   & 356.69968 & 2.46240 & 20.01 & 0.3343 & 9.04 & 1.18 & 1.01 & 8.15 \\
    \enddata
    \tablecomments{(5): spectroscopic redshifts measured using the redshifted [O\textsc{iii}]$_{5007}$ emission lines;
    (6): stellar mass from the SED fitting (Section~\ref{sec:sed});
    (7): SFR from the SED fitting (Section~\ref{sec:sed});
    (8): SFR from the extinction-corrected H$\alpha$ line flux \citep[using the method from][]{1998ARAA..36..189K};
    (9): oxygen abundance based on the $N2$ method (Section~\ref{sec:metallicity})}
\end{deluxetable*}

\begin{deluxetable*}{lccccccccc}
    \tabletypesize{\scriptsize}
    \tablecaption{\label{tab:lineflux} Emission line flux measurements in 26 DES-GPs}
    \tablehead{
    \colhead{GP ID} & \colhead{H$\alpha$} & \colhead{H$\beta$} & \colhead{[O\textsc{ii}]$_{3727,3729}$} &
    \colhead{[O\textsc{iii}]$_{4959}$} & \colhead{[O\textsc{iii}]$_{5007}$} &
    \colhead{[N\textsc{ii}]$_{6584}$} &
    \colhead{[S\textsc{ii}]$_{6717}$} & \colhead{[S\textsc{ii}]$_{6731}$}
    }
    \startdata
    GP000207$-$424024 & $52.40\pm0.19$ & $15.87\pm0.20$ & $24.59\pm0.90$ & $21.19\pm0.22$ & $70.81\pm0.35$ & $3.20\pm0.08$ & $1.09\pm0.09$ & $1.72\pm0.14$ \\
    GP000957+024146   & $44.06\pm0.20$ & $11.21\pm0.30$ & $11.30\pm1.50$ & $17.36\pm0.31$ & $50.19\pm0.48$ & $3.19\pm0.11$ & $2.81\pm0.12$ & $2.32\pm0.12$ \\
    GP001034$-$484901 & $68.54\pm0.26$ & $14.06\pm0.37$ & $16.10\pm1.90$ & $28.02\pm0.45$ & $67.35\pm0.48$ & $1.97\pm0.12$ & $3.72\pm0.18$ & $2.20\pm0.13$ \\
    GP005423+033009   & $23.60\pm0.25$ & $5.79\pm0.36$ & $7.60\pm1.80$ & $6.87\pm0.38$ & $24.22\pm0.37$ & $2.00\pm0.15$ & $1.78\pm0.12$ & $1.23\pm0.15$ \\
    GP012256$-$154025 & $63.30\pm0.36$ & $18.10\pm0.38$ & $32.70\pm3.00$ & $34.50\pm0.41$ & $99.30\pm0.50$ & $2.95\pm0.19$ & $4.98\pm0.20$ & $3.15\pm0.16$ \\
    GP022429$-$105748 & $57.80\pm0.31$ & $15.40\pm0.29$ & $\cdots$ & $32.50\pm0.32$ & $103.00\pm0.47$ & $2.32\pm0.16$ & $3.37\pm0.19$ & $2.55\pm0.11$ \\
    GP024352+035504   & $34.80\pm0.17$ & $9.54\pm0.20$ & $16.40\pm1.30$ & $16.10\pm0.21$ & $55.10\pm0.28$ & $1.95\pm0.11$ & $1.99\pm0.16$ & $1.55\pm0.09$ \\
    GP031823$-$412811 & $65.90\pm0.23$ & $18.20\pm0.26$ & $31.60\pm1.30$ & $30.40\pm0.28$ & $95.70\pm0.37$ & $3.77\pm0.11$ & $2.71\pm0.20$ & $3.09\pm0.20$ \\
    GP033528$-$562307 & $16.00\pm0.12$ & $4.85\pm0.23$ & $7.60\pm1.30$ & $9.32\pm0.15$ & $27.90\pm0.23$ & $0.56\pm0.07$ & $0.95\pm0.07$ & $0.71\pm0.13$ \\
    GP035844$-$272134 & $130.00\pm0.37$ & $38.80\pm0.36$ & $46.80\pm2.40$ & $78.40\pm0.46$ & $237.00\pm0.73$ & $3.92\pm0.13$ & $5.89\pm0.12$ & $4.41\pm0.13$ \\
    GP042045$-$600735 & $49.00\pm0.29$ & $13.80\pm0.29$ & $18.30\pm2.20$ & $27.50\pm0.31$ & $81.20\pm0.42$ & $2.17\pm0.17$ & $2.32\pm0.11$ & $1.97\pm0.11$ \\
    GP043224$-$383800 & $45.30\pm0.30$ & $11.70\pm0.35$ & $19.00\pm2.70$ & $20.60\pm0.37$ & $60.90\pm0.41$ & $2.41\pm0.13$ & $3.97\pm0.16$ & $3.08\pm0.22$ \\
    GP051716$-$265543 & $68.10\pm0.28$ & $18.70\pm0.31$ & $21.60\pm2.10$ & $37.70\pm0.45$ & $114.00\pm0.47$ & $2.22\pm0.22$ & $2.42\pm0.12$ & $1.82\pm0.11$ \\
    GP052936$-$362507 & $65.50\pm0.24$ & $19.10\pm0.35$ & $23.80\pm2.40$ & $37.50\pm0.36$ & $112.00\pm0.49$ & $1.76\pm0.11$ & $3.64\pm0.22$ & $2.55\pm0.15$ \\
    GP054214$-$354139 & $35.10\pm0.24$ & $9.53\pm0.26$ & $20.80\pm1.90$ & $16.00\pm0.27$ & $49.20\pm0.34$ & $2.33\pm0.18$ & $2.93\pm0.13$ & $2.08\pm0.12$ \\
    GP204045$-$413555 & $26.19\pm0.09$ & $8.07\pm0.12$ & $\cdots$ & $15.71\pm0.14$ & $46.17\pm0.17$ & $0.65\pm0.04$ & $0.49\pm0.09$ & $0.31\pm0.11$ \\ 
    GP204359$-$540359 & $41.20\pm0.16$ & $9.11\pm0.17$ & $17.10\pm0.93$ & $20.36\pm0.18$ & $61.66\pm0.24$ & $2.34\pm0.11$ & $1.86\pm0.10$ & $1.45\pm0.13$ \\
    GP210903$-$444951 & $20.90\pm0.18$ & $6.07\pm0.36$ & $\cdots$ & $11.18\pm0.31$ & $34.79\pm0.35$ & $0.72\pm0.16$ & $0.98\pm0.11$ & $0.69\pm0.10$ \\
    GP211040$-$504353 & $40.48\pm0.20$ & $11.32\pm0.27$ & $9.60\pm1.30$ & $25.18\pm0.28$ & $73.95\pm0.36$ & $1.49\pm0.16$ & $1.58\pm0.09$ & $1.37\pm0.13$ \\
    GP211619$-$463914 & $12.25\pm0.18$ & $4.56\pm0.23$ & $5.18\pm0.66$ & $9.35\pm0.22$ & $31.19\pm0.29$ & $0.77\pm0.10$ & $0.57\pm0.10$ & $0.47\pm0.11$ \\
    GP212733$-$424114 & $16.98\pm0.10$ & $5.16\pm0.16$ & $5.26\pm0.51$ & $8.87\pm0.16$ & $26.39\pm0.18$ & $0.84\pm0.09$ & $0.90\pm0.11$ & $0.93\pm0.09$ \\
    GP213914$-$495028 & $23.10\pm0.18$ & $7.11\pm0.21$ & $7.76\pm0.86$ & $13.81\pm0.21$ & $40.78\pm0.30$ & $1.52\pm0.11$ & $0.94\pm0.11$ & $0.50\pm0.12$ \\
    GP215810$-$615644 & $19.61\pm0.18$ & $4.72\pm0.20$ & $\cdots$ & $8.58\pm0.21$ & $26.02\pm0.32$ & $1.09\pm0.09$ & $1.59\pm0.08$ & $1.13\pm0.08$ \\
    GP232739$-$454554 & $17.23\pm0.14$ & $0.94\pm0.15$ & $\cdots$ & $2.18\pm0.17$ & $8.56\pm0.22$ & $0.78\pm0.08$ & $\cdots$ & $0.86\pm0.10$ \\
    GP234638$-$011839 & $43.70\pm0.26$ & $13.00\pm0.41$ & $24.00\pm1.90$ & $22.27\pm0.37$ & $68.20\pm0.47$ & $2.12\pm0.20$ & $2.36\pm0.14$ & $1.85\pm0.12$ \\
    GP234647+022744   & $28.42\pm0.20$ & $9.31\pm0.27$ & $\cdots$ & $16.98\pm0.26$ & $46.04\pm0.39$ & $1.39\pm0.15$ & $1.40\pm0.13$ & $0.94\pm0.09$ \\
    \enddata
    \tablecomments{Line fluxes are in units of $10^{-16}\mathrm{erg}\,  \mathrm{s}^{-1}\,\mathrm{cm}^{-2}$.}
\end{deluxetable*}

\begin{deluxetable*}{lcccccccc}
    \tabletypesize{\scriptsize}
    \tablecaption{\label{tab:ew} Observed equivalent width measurements in 26 DES-GPs}
    \tablehead{
    \colhead{GP ID} & \colhead{H$\alpha$} & \colhead{H$\beta$} & \colhead{[O\textsc{ii}]$_{3727,3729}$} &
    \colhead{[O\textsc{iii}]$_{4959}$} & \colhead{[O\textsc{iii}]$_{5007}$} &
    \colhead{[N\textsc{ii}]$_{6584}$} &
    \colhead{[S\textsc{ii}]$_{6717}$} & \colhead{[S\textsc{ii}]$_{6731}$}
    }
    \startdata
    GP000207$-$424024 & $260\pm7$ & $91\pm6$ & $67\pm13$ & $120\pm7$ & $421\pm15$ & $14\pm2$ & $14\pm4$ & $22\pm9$ \\
    GP000957+024146   & $442\pm11$ & $63\pm8$ & $27\pm15$ & $91\pm8$ & $223\pm11$ & $31\pm5$ & $33\pm7$ & $35\pm8$ \\
    GP001034$-$484901 & $422\pm9$ & $49\pm5$ & $33\pm15$ & $112\pm9$ & $267\pm10$ & $12\pm3$ & $36\pm8$ & $23\pm6$ \\
    GP005423+033009   & $199\pm12$ & $23\pm7$ & $8\pm8$ & $24\pm6$ & $89\pm7$ & $19\pm6$ & $15\pm5$ & $11\pm6$ \\
    GP012256$-$154025 & $452\pm14$ & $61\pm6$ & $28\pm10$ & $120\pm7$ & $323\pm8$ & $20\pm5$ & $56\pm9$ & $22\pm5$ \\
    GP022429$-$105748 & $671\pm20$ & $96\pm8$ & $\cdots$ & $175\pm7$ & $543\pm12$ & $29\pm9$ & $43\pm11$ & $33\pm6$ \\
    GP024352+035504   & $418\pm12$ & $58\pm6$ & $38\pm13$ & $98\pm7$ & $313\pm9$ & $24\pm6$ & $27\pm9$ & $21\pm5$ \\
    GP031823$-$412811 & $598\pm13$ & $78\pm5$ & $47\pm9$ & $131\pm6$ & $372\pm8$ & $32\pm4$ & $22\pm7$ & $26\pm7$ \\
    GP033528$-$562307 & $357\pm15$ & $50\pm10$ & $21\pm14$ & $74\pm5$ & $238\pm10$ & $13\pm7$ & $23\pm8$ & $18\pm14$ \\
    GP035844$-$272134 & $586\pm9$ & $120\pm5$ & $50\pm12$ & $248\pm8$ & $755\pm13$ & $20\pm2$ & $39\pm3$ & $30\pm3$ \\
    GP042045$-$600735 & $457\pm16$ & $73\pm7$ & $19\pm10$ & $121\pm7$ & $372\pm10$ & $22\pm8$ & $21\pm4$ & $21\pm5$ \\
    GP043224$-$383800 & $390\pm13$ & $52\pm7$ & $17\pm10$ & $80\pm7$ & $233\pm8$ & $23\pm5$ & $40\pm7$ & $31\pm12$ \\
    GP051716$-$265543 & $796\pm21$ & $87\pm7$ & $26\pm12$ & $141\pm8$ & $470\pm11$ & $25\pm11$ & $25\pm6$ & $20\pm5$ \\
    GP052936$-$362507 & $435\pm9$ & $75\pm7$ & $22\pm8$ & $144\pm7$ & $391\pm10$ & $12\pm3$ & $30\pm8$ & $21\pm5$ \\
    GP054214$-$354139 & $371\pm13$ & $50\pm6$ & $22\pm8$ & $76\pm6$ & $233\pm8$ & $26\pm7$ & $32\pm7$ & $23\pm5$ \\
    GP204045$-$413555 & $455\pm9$ & $76\pm5$ & $\cdots$ & $159\pm7$ & $467\pm9$ & $11\pm2$ & $22\pm12$ & $12\pm16$ \\
    GP204359$-$540359 & $414\pm10$ & $75\pm7$ & $41\pm11$ & $174\pm8$ & $520\pm14$ & $30\pm8$ & $24\pm6$ & $18\pm7$ \\
    GP210903$-$444951 & $246\pm11$ & $33\pm9$ & $\cdots$ & $59\pm7$ & $193\pm9$ & $7\pm6$ & $10\pm4$ & $9\pm5$ \\
    GP211040$-$504353 & $405\pm11$ & $63\pm6$ & $13\pm6$ & $155\pm7$ & $332\pm7$ & $17\pm7$ & $27\pm6$ & $18\pm6$ \\
    GP211619$-$463914 & $127\pm26$ & $46\pm10$ & $22\pm10$ & $111\pm11$ & $319\pm12$ & $52\pm32$ & $16\pm9$ & $14\pm11$ \\
    GP212733$-$424114 & $496\pm14$ & $66\pm9$ & $27\pm11$ & $106\pm8$ & $325\pm10$ & $24\pm11$ & $29\pm12$ & $29\pm10$ \\
    GP213914$-$495028 & $492\pm24$ & $59\pm8$ & $22\pm11$ & $132\pm9$ & $308\pm12$ & $33\pm11$ & $27\pm12$ & $9\pm6$ \\
    GP215810$-$615644 & $446\pm25$ & $38\pm7$ & $\cdots$ & $72\pm8$ & $219\pm13$ & $26\pm10$ & $70\pm18$ & $106\pm29$ \\
    GP232739$-$454554 & $334\pm16$ & $8\pm5$ & $\cdots$ & $22\pm7$ & $95\pm12$ & $16\pm7$ & $\cdots$ & $19\pm9$ \\
    GP234638$-$011839 & $298\pm11$ & $43\pm8$ & $37\pm13$ & $66\pm6$ & $227\pm10$ & $18\pm7$ & $19\pm6$ & $15\pm4$ \\
    GP234647+022744   & $471\pm19$ & $70\pm10$ & $\cdots$ & $127\pm10$ & $356\pm16$ & $26\pm14$ & $31\pm13$ & $21\pm10$ \\
    \enddata
    \tablecomments{Equivalent widths are in units of $\mathrm{\AA}$.}
\end{deluxetable*}

%%% ACKNOWLEDGMENTS (IF ANY) %%%%%%%%%%%%%%%%%%%%%%%%%%%%%%%%%%%%%%%%

\begin{acknowledgments}
This work was supported by the National Research Foundation of Korea (NRF) grant funded by the Korea government (MSIT) (Nos. 2022R1A4A3031306, RS-2024-00349364). J.H.L. acknowledges support from Basic Science Research Program through the National Research Foundation of Korea (NRF) funded by the Ministry of Education (No. RS-2024-00452816). This work was supported by K-GMT Science Program (PID: GS-2024A-Q-313, GS-2024A-Q-411, GS-2024B-Q-313) of Korea Astronomy and Space Science Institute (KASI). Section~\ref{sec:spectro} and following results are based on observations obtained 
at the international Gemini Observatory, a program of NSF NOIRLab, which is managed by the Association of Universities for Research in Astronomy (AURA) under a cooperative agreement with the U.S. National Science Foundation on behalf of the Gemini Observatory partnership: the U.S. National Science Foundation (United States), National Research Council (Canada), Agencia Nacional de Investigaci\'{o}n y Desarrollo (Chile), Ministerio de Ciencia, Tecnolog\'{i}a e Innovaci\'{o}n (Argentina), Minist\'{e}rio da Ci\^{e}ncia, Tecnologia, Inova\c{c}\~{o}es e Comunica\c{c}\~{o}es (Brazil), and Korea Astronomy and Space Science Institute (Republic of Korea). 
\end{acknowledgments}


\vspace{5mm}
\facility{Gemini:South}


\software{astropy \citep{2013A&A...558A..33A,2018AJ....156..123A},  
          jdaviz \citep{2022zndo...5513927D}
          %Cloudy \citep{2013RMxAA..49..137F}, 
          %Source Extractor \citep{1996A&AS..117..393B}
          }




\bibliography{gp}{}
\bibliographystyle{aasjournal}



\end{document}